\documentclass{ewass_ss4proc}

\editors{Emeline Bolmont \& Sergi Blanco-Cuaresma}
\editorsurl{www.emelinebolmont.com | www.blancocuaresma.com/s/}
\publisher{Zenodo}
\conference{EWASS Special Session 4 (2017): Star-planet interactions}
\conferencedate{2017}

\title{Studying tidal effects in planetary systems with Posidonius. \\ A N-body simulator written in Rust. }
\author{Sergi Blanco-Cuaresma$^{1,2}$ 
        Emeline Bolmont$^{3}$}

\affiliation{$^{1}$ Harvard-Smithsonian Center for Astrophysics, 60 Garden Street, Cambridge, MA 02138, USA  \\
			 $^{2}$ Observatoire de Gen\`eve, Universit\'e de Gen\`eve, CH-1290 Versoix, Switzerland \\
			 $^{3}$ Laboratoire AIM Paris-Saclay, CEA/DRF-CNRS-Univ. Paris Diderot - IRFU/SAp, Centre de Saclay, F- 91191 Gif-sur-Yvette Cedex, France}

\shorttitle{Posidonius}
\shortauthors{Sergi Blanco-Cuaresma \& Emeline Bolmont}

\abs{Planetary systems with several planets in compact orbital configurations such as TRAPPIST-1 are surely affected by tidal effects. Its study provides us with important insight about its evolution. We developed a second generation of a N-body code based on the tidal model used in \mbox{Mercury-T}, re-implementing and improving its functionalities using Rust as programming language (including a Python interface for easy use) and the WHFAST integrator. The new open source code ensures memory safety, reproducibility of numerical N-body experiments, it improves the spin integration compared to \mbox{Mercury-T} and allows to take into account a new prescription for the dissipation of tidal inertial waves in the convective envelope of stars. Posidonius is also suitable for binary system simulations with evolving stars.}

\begin{document}

\maketitle

\section{Introduction}

Detection of planetary systems around ultracool dwarfs (i.e., stars with effective temperatures lower than 2\,700 K) are expected to increase in the coming years \citep{2016Natur.533..221G} thanks to on-going efforts such as the TRAPPIST survey \citep{2011Msngr.145....2J} and future missions such as the ambitious project SPECULOOS (Search for habitable Planets EClipsing ULtra-cOOl Stars) that will be installed at ESO's Paranal Observatory.  These systems can have several planets in compact orbital configurations, and they can be in or close to mean motion resonances as already observed in TRAPPIST-1 \citep{2017Natur.542..456G}.  Planets arranged in this kind of set-up are surely affected by tidal effects, and the use of N-body simulation is necessary to account for the complicated dynamics of the system.

One of the most used open source tools by the community to simulate exoplanetary systems is \mbox{Mercury-T}\footnote{http://www.emelinebolmont.com/} \citep{2015A&A...583A.116B}, which is based on the N-body simulator Mercury \citep{1999MNRAS.304..793C} with an added extension to account for tidal effects, general relativity, rotational flattening and the evolution of the central body. The original Mercury is a customizable general-purpose N-body code written in Fortran 77, although there are forks re-written in Fortran 90\footnote{https://code.google.com/archive/p/mercury-90/} that were actually used to create \mbox{Mercury-T}. 

Fortran (derived from Formula Translation) is an imperative programming language created by IBM in the 1950s and it is especially suited to scientific computing. The language has gone through several updates since its birth (where version names correlate with the year the update was released), adding support for structured programming (Fortran 77), modular and generic programming (Fortran 90), high performance capabilities (Fortran 95), object-oriented programming (Fortran 2003) and concurrent programming (Fortran 2008). Nevertheless, most of the scientific community keeps using Fortran 77/90 and it is common than available codes only work properly with specific compilers (e.g., the propietary Intel Fortran/ifort Compiler, the deprecated g77 or the modern GNU Fortran/gfortran). Fortran (like the C language) provides to the developer a high level of control for managing the computer memory which can lead to errors and bugs. For instance, Mercury suffered a problem due to an array not being properly initialized, hence its elements may access random values which will affect the simulation results \citep{2008arXiv0808.0483D} and the scientific conclusions.

\begin{figure*}[!htb]
\center
\includegraphics[width=0.3\linewidth]{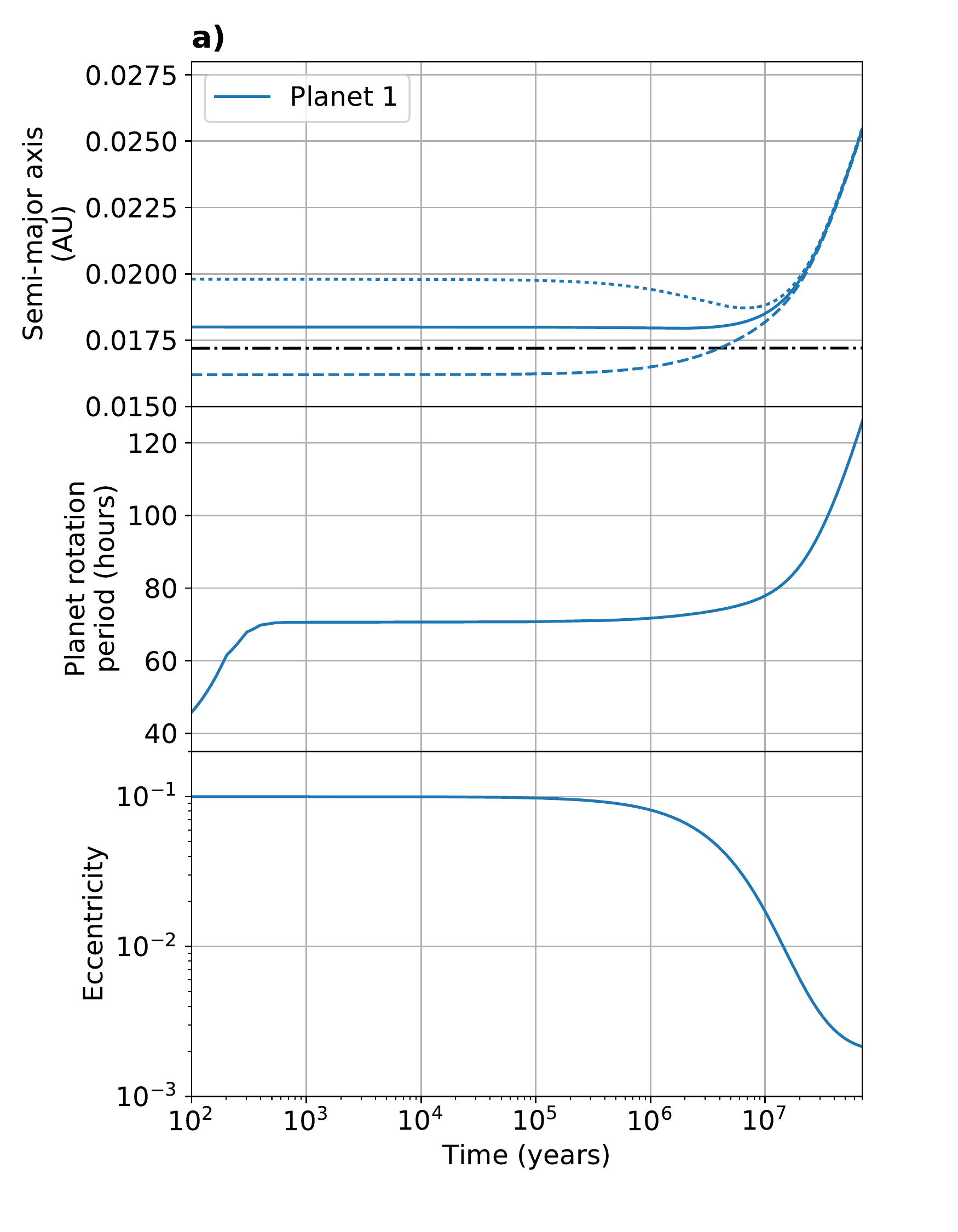}
\includegraphics[width=0.3\linewidth]{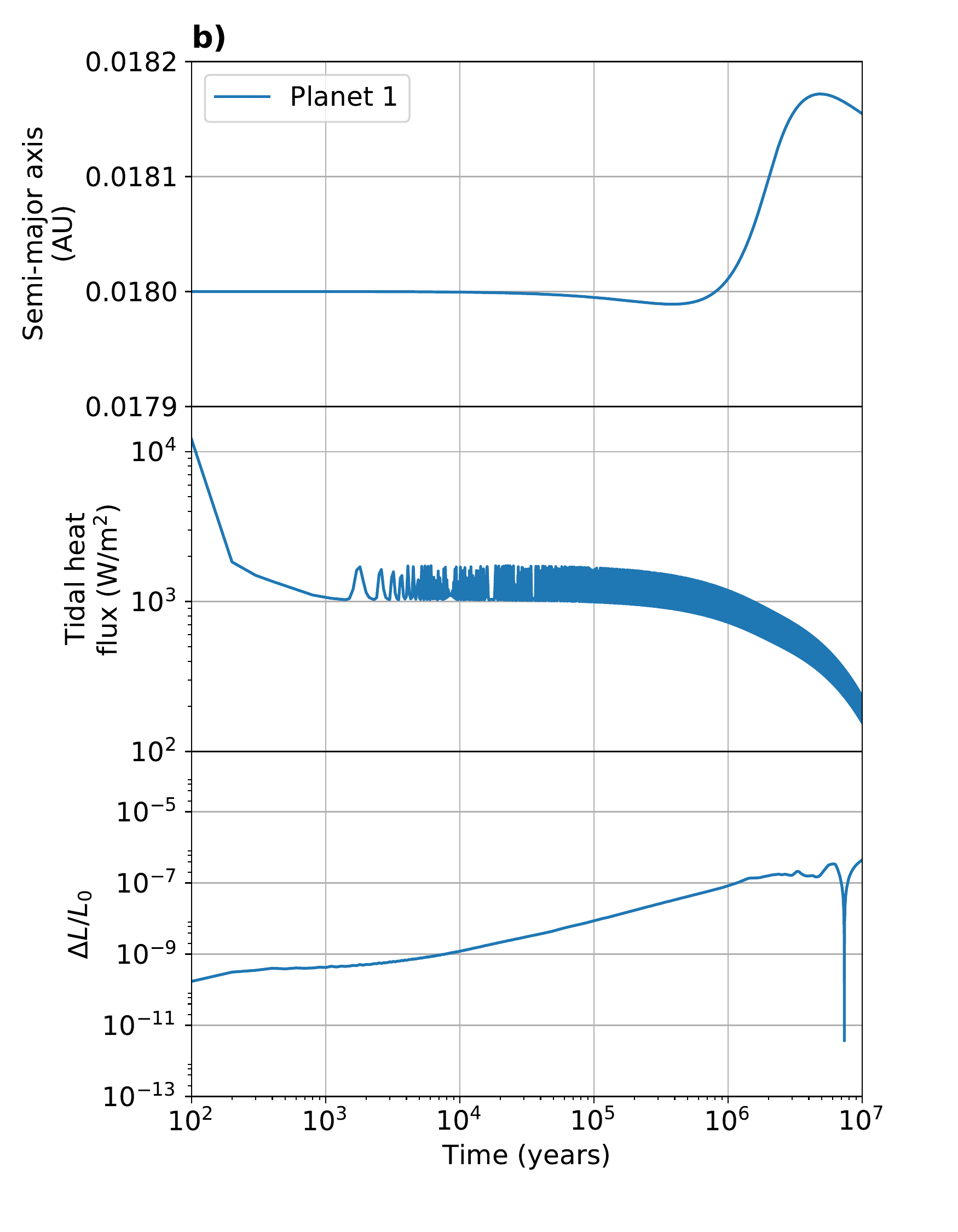}
\includegraphics[width=0.3\linewidth]{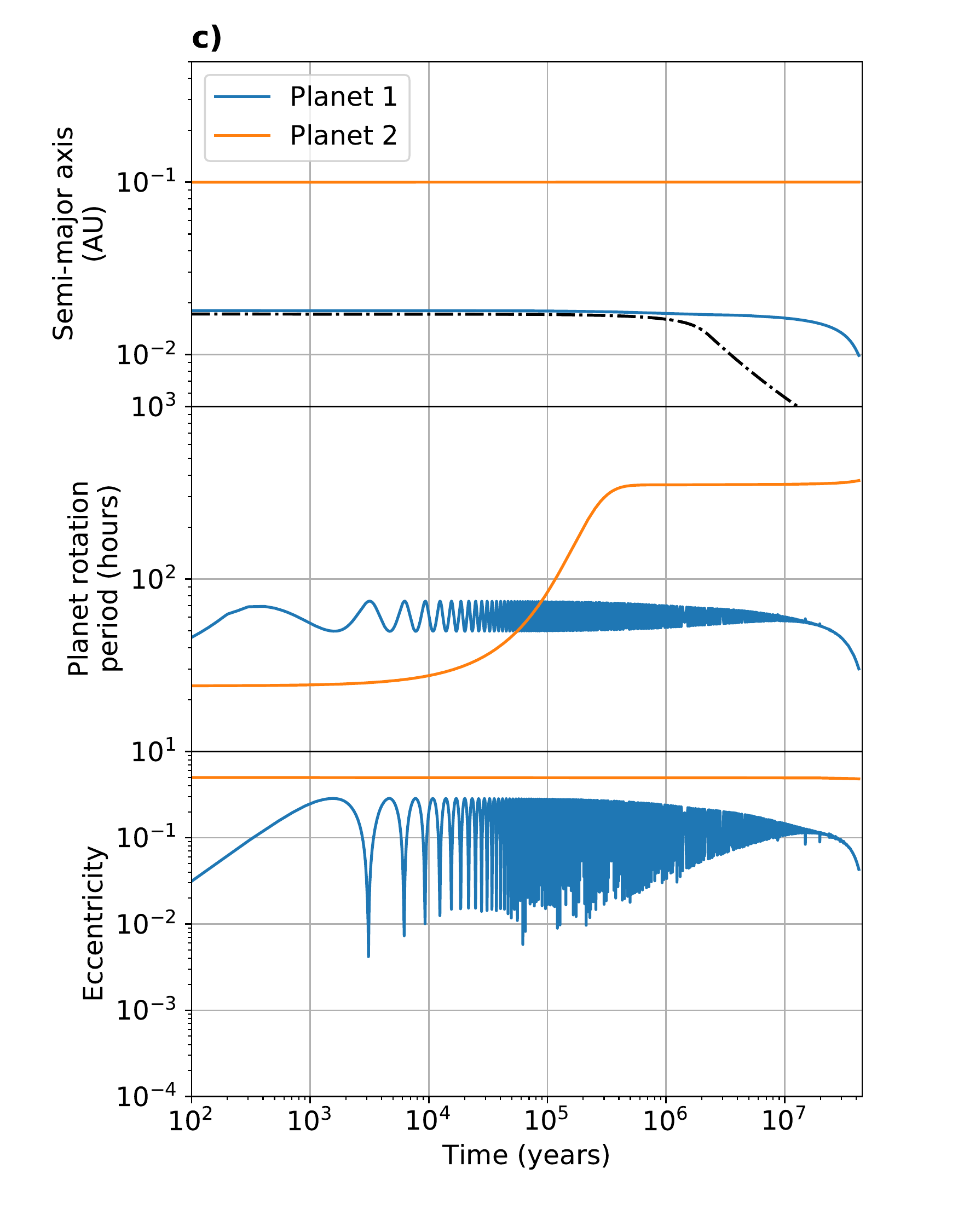}
\caption{\label{fig:case34527} a) Case 3 from \cite{2015A&A...583A.116B} where the dashed blue lines correspond to the periastron and apostron and the black dashed line is the co-rotation radius in the upper figure. b) Case 4 from \cite{2015A&A...583A.116B} where the conservation of angular momentum is shown in the lower figure. c) Case exposed in Sect.~5.2.2 from \cite{bolmo2013} (Fig.~5.27) where the black dashed line is the co-rotation radius in the upper figure.}
\end{figure*}

Certain Fortran compilers offer validations such as checking for initialized variables or checking out-of-bound access for arrays, which can be enabled/disabled depending on the flags used to compile the code. These verifications, although useful, do not cover all possible problems derived from memory management and, in any case, they are not enforced by the language. Given the wide use of the language in the scientific community, the lack of software engineering training in certain disciplines such as astrophysics and the lack of memory management controls in the language, it can be convenient to look for alternative technologies. A possibility could be to use garbage collected languages such as Python or Java, which are already extensively used by the astronomy community (especially Python). But these solutions have an impact on execution times, and codes such as N-body simulators are computationally intense and they require fast solutions to be useful. It is for these specific cases that the new programming language Rust can be well fitted as exposed in \cite{2017IAUS..325..341B}.

We developed Posidonius\footnote{http://www.blancocuaresma.com/s/} \footnote{Posidonius (c. 135 BCE - c. 51 BCE), was a Greek Stoic philosopher, politician, astronomer, geographer, historian and teacher native to Apamea (Syria). In Hispania, on the Atlantic coast near Cadiz, Posidonius could observe tides much higher than in his native Mediterranean. He wrote that daily tides are related to the Moon's orbit, while tidal heights vary with the cycles of the Moon.} and distributed it under an open source license\footnote{GNU Affero General Public License}, a second generation of N-body code based on the tidal model used in \mbox{Mercury-T} \citep{2015A&A...583A.116B}, re-implementing and improving its functionalities using Rust as programming language and the WHFAST integrator \citep{2015MNRAS.452..376R}. The new code ensures memory safety, reproducibility of numerical N-body experiments, it improves the spin integration compared to \mbox{Mercury-T} and allows to take into account a new prescription for the dissipation of tidal inertial waves in the convective envelope of stars \citep{2016CeMDA.126..275B, 2017A&A...604A.112G, 2017A&A...604A.113B}.

\section{The design}

Posidonius is composed by a N-body simulator written in Rust and a Python package used to define the initial condition of the scientific cases the user wants to simulate. Rust ensures that the simulation is free from typical memory management errors but its learning curve can be steep \citep{2017IAUS..325..341B}, hence with this separation, we provide an user-friendly scripting interface and do not force the user to learn Rust details unless a more advanced use is really necessary (e.g., implementing new physical effects, developing a new integrator). Once the case is defined, the python script generates a JSON format file describing the initial scenario (actually, with a bit of effort, that JSON file could be generated manually or using any other programming language) and Posidonius can read this file to initiate the simulation.

An important aspect of N-body simulations (and more in general, for any scientific study) is ensuring that your computations can be reproduced and reverified \citep{2017MNRAS.467.2377R}. Posidonius stores the history of the simulation together with frequent recovery snapshots in binary format. The user can stop and resume any simulation at any point or repeat again the same calculations by using an older recovery snapshot. But, more importantly, any study can be completely reproduced by indicating what Posidonius version was used and providing the original JSON files with the initial conditions (or the Python script that generates it).

From a technical point of view, we have taken some decisions to optimize the execution time. We do not use certain Rust capabilities such as vectors (i.e., re-sizable arrays allocated in the heap), all the bodies of a simulation are stored in arrays allocated in the stack for faster access. The array has a fixed size for all the simulations (defined by a constant in the code), it is possible to create cases with a lower number of bodies but not higher (although that can be overcome by changing the constant value and recompiling Posidonius). We do not take full advantage of the possibilities of Rust for memory management due to this decision, but we are open to study and incorporate alternative technical designs that other experts may come up with.

To compute the evolution of positions and velocities from gravity forces, Posidonius implements the symplectic integrator WHFAST \citep{2015MNRAS.452..376R} without correctors (since tidal forces are velocity dependent). This algorithm has been proved to be fast and accurate for long-term orbit integrations of planetary systems. The code can execute integrations using Jacobi, Democratic-Heliocentric or WHDS coordinates \citep{2017MNRAS.468.2614H} as provided in the REBOUND N-body code \citep{2012A&A...537A.128R}. In a consistent way, the spin evolution is computed for all the bodies by using a midpoint integrator \citep{Stoer2013}.

Tidal forces, rotational-flattening effects and general relativity corrections are implemented following the model described by \cite{2015A&A...583A.116B}. In the case of the general relativity, we also included the prescriptions from \cite{1975ApJ...200..221A} and \cite{1983A&A...125..150N} (the latter considers effects from all bodies in the system, making Posidonius ready for binary stars too) following the REBOUND implementation. Additionally, evolution models for FGKML stars and gaseous planets have been included. Some of these models were already present in Mercury-T: brown dwarfs models of \cite{2011A&A...528A..41L}, 0.1 and 1~$M_\odot$ models of \cite{1998A&A...337..403B} Jupiter model of \cite{2013NatGe...6..347L}. In Posidonius, we added the evolutionary models used to compute tidal evolution in \cite{2016CeMDA.126..275B}. These models are STAREVOL models for FGKM stars \citep{2000A&A...358..593S}. We also implemented stellar evolution models, coming from a more recent version of the STAREVOL code \citep{2017A&A...604A.112G}.  All these evolution models give the evolution of stellar quantities such as the radius of the star with time. The newest models \citep{2017A&A...604A.112G} allow to take into account the evolution of the frequency-averaged dissipation in the convective envelope of FGKM stars.
For these last models, the user must keep in mind that the physical validity of the dissipation formulation is for coplanar, circular orbits, which is never the case if there are more than one planet in the system.

\section{The assessment}

The best approach to assess the quality of Posidonius is to repeat simulations that were previously executed, peer-reviewed and published based on \mbox{Mercury-T}. We have selected the case three and four defined in Table~4 from \cite{2015A&A...583A.116B} (non-evolving and evolving brown dwarf with one planet and considering tidal effects), case six to six triple prime listed in Table~5 from \cite{2015A&A...583A.116B} (non-evolving brown dwarf with two planets and considering rotational-fattening using time steps of 0.08, 0.05, 0.01, 0.001 days) and the case investigated in Sect.~5.2.2 from \cite{bolmo2013} (evolving brown dwarf with two planets considering tidal effects and general relativity corrections). To facilitate the reproducibility of any of these tests, the initial condition for each of them are included in the Posidonius source code. 

\begin{figure}[!htb]
\center
\includegraphics[width=0.7\linewidth]{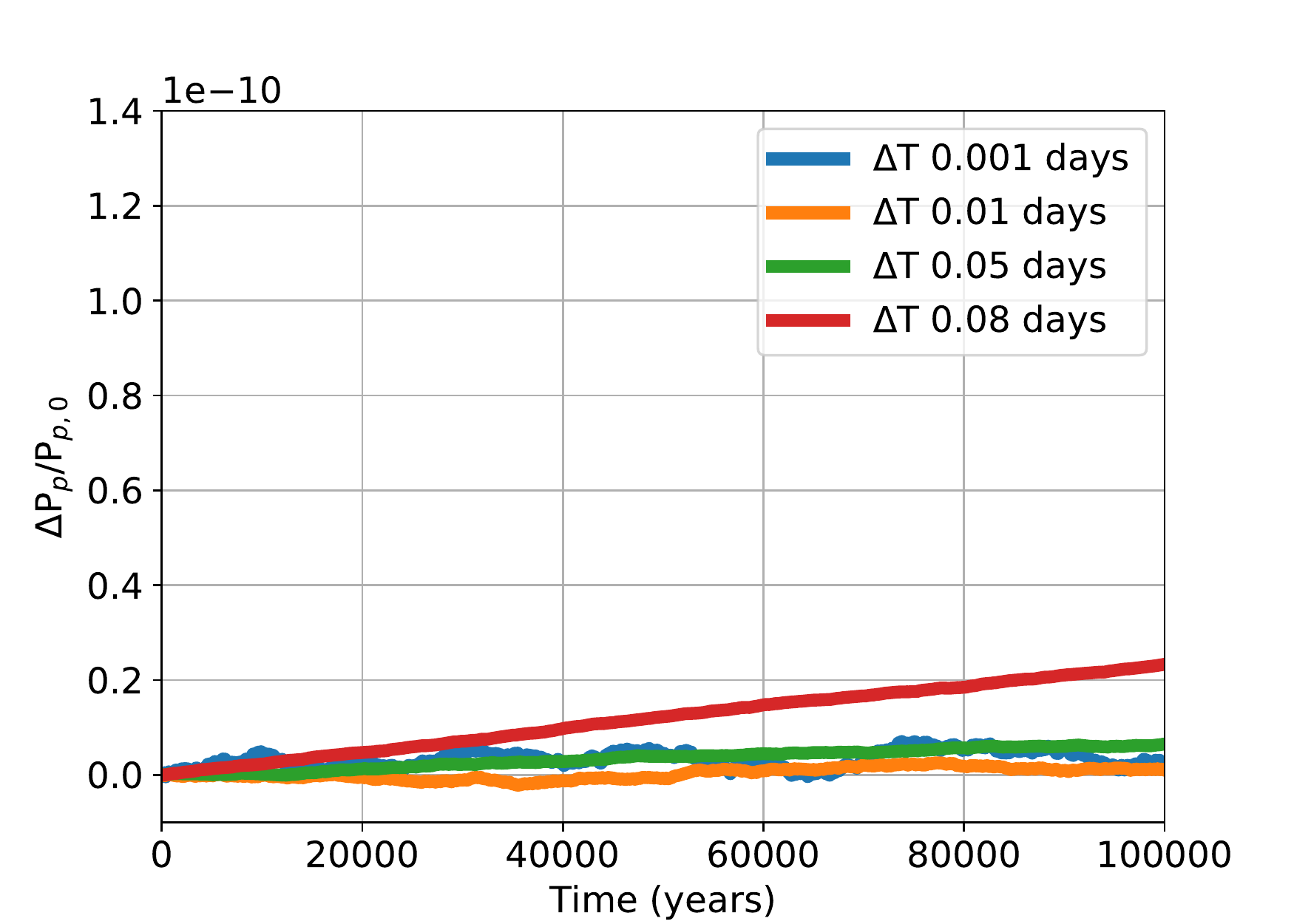}
\caption{\label{fig:case6} Conservation of planet rotation period (at $10^{-10}$ scale) for different time steps, equivalent to Fig.~7 (case 6) from \cite{2015A&A...583A.116B}. Difference in rotation period constantly increases for time steps equal or greater than 0.05, for lower values the difference follows a random walk centered around zero when considering longer time spans.
}
\end{figure}

The resulting semi-major axis evolution for cases three and four are shown in Fig.~\ref{fig:case34527}a and Fig.~\ref{fig:case34527}b and they match the results presented in Fig.~3 and Fig.~6 from \cite{2015A&A...583A.116B}. We observed that the conservation of angular momentum for case four (\ref{fig:case34527}b) is one order of magnitude better than \mbox{Mercury-T}. On the problem of \mbox{Mercury-T} having a numerical shift of the rotation period of the planets when only the rotational flattening effect was on (cases six to six triple prime), Posidonius has a much better behavior (five orders of magnitude better when Fig.~\ref{fig:case6} is compared to Fig.~7 from \cite{2015A&A...583A.116B}). Finally, the resulting evolution for the last case with multiple planets is shown in Fig.~\ref{fig:case34527} and it also very similar to Fig.~5.27 from \cite{bolmo2013}. Although we do not include the results in this work due to space limitations, it is worth mentioning that we successfully reproduced the results of Fig.~5.28 and Fig.~5.30 from \cite{bolmo2013}, which explore the effects of the stellar dissipation factor and resonances.

\begin{figure}[!htb]
\center
\includegraphics[width=0.7\linewidth]{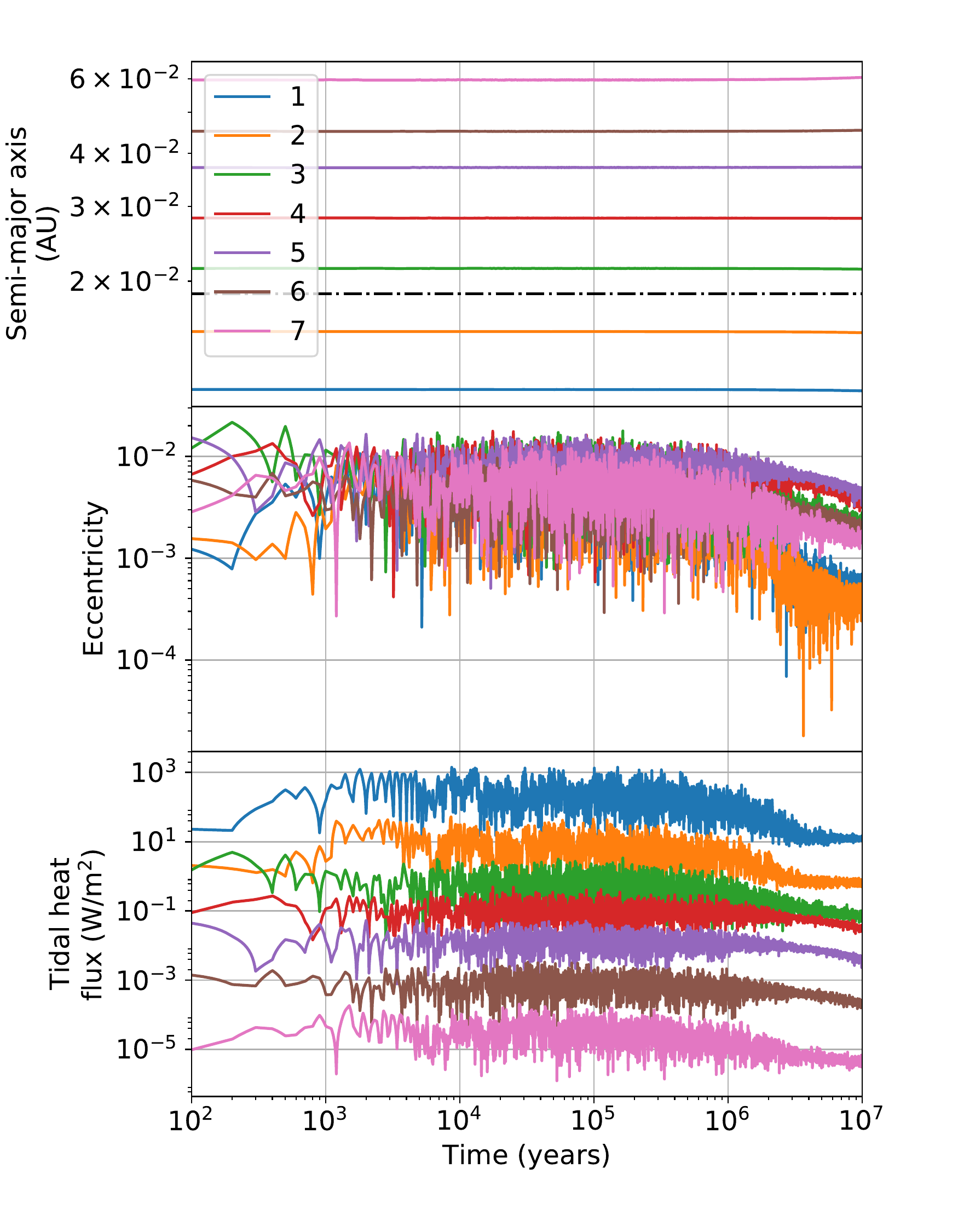}
\caption{\label{fig:trappist1} TRAPPIST-1 simulation with parameters from Table~\ref{tab:trappist1}. The black dashed line in the upper figure corresponds to the co-rotation radius.
}
\end{figure}

In parallel, during the execution of these tests, we measured and compared Posidonius speed and we observed that it is six times faster than \mbox{Mercury-T} in average. This is a significant improvement that is not strictly linked to the chosen programming language, the new code was written with different design patterns not followed in \mbox{Mercury-T}.

\begin{table*}[!htb]
	\centering
	\caption{TRAPPIST-1 planetary initial conditions used in the simulation shown in Fig.~\ref{fig:trappist1}.}
	\label{tab:trappist1}
	\begin{tabular*}{0.85\linewidth}{c @{\extracolsep{\fill}} c c c c c}
	\noalign{\smallskip}\hline\hline\noalign{\smallskip}
	Planet & Radius & Mass & Semi-major axis & Inclination & Mean anomaly \\
	& (R$_{\bigoplus}$) & ($M_{\odot}$) & (AU) & (deg) & (deg) \\
	\noalign{\smallskip}\hline\noalign{\smallskip}
    1 &  1.086 & 2.5843686$\times 10^{-6}$ & 0.01111 & 0.35 & 323.732652895  \\
	2 &  1.056 & 4.1957984$\times 10^{-6}$ & 0.01521 & 0.33 & 96.4925777097  \\
	3 &  0.772 & 1.2465778$\times 10^{-6}$ & 0.02144 & 0.25 & 111.770368348  \\
	4 &  0.918 & 1.8850689$\times 10^{-6}$ & 0.02817 & 0.14 & 165.724187804  \\
	5 &  1.045 & 2.0674949$\times 10^{-6}$ & 0.03710 & 0.32 & 254.117367005  \\
	6 &  1.127 & 4.0741811$\times 10^{-6}$ & 0.04510 & 0.29 & 161.020362506  \\
	7 &  0.755 & 1.2465778$\times 10^{-6}$ & 0.05960 & 0.13 & 134.724813585  \\
	\noalign{\smallskip}\hline
	\end{tabular*}
\end{table*}

\section{The scientific case: TRAPPIST-1}

We chose the TRAPPIST-1 system as an example of a useful scientific case to be studied with Posidonius. We run a simulation with a duration of $10^{7}$ years with a time step of 0.08 days, where all the effects where enabled (i.e, tides, rotational flattening and general relativity corrections). The star (just above the brown dwarf limit for this system) was set to have a mass of 0.08 M$_{\odot}$, a radius of 0.117 R$_{\odot}$, a rotation period of 3.3 hours, a low dissipation factor (1\% of a typical brown dwarf, \citealt{2011A&A...535A..94B}), and radius of gyration and love number similar to brown dwarfs. No stellar evolution was enabled because TRAPPIST-1 host star is in an old and stable phase of its life. The parameters for the seven planets in the system are obtained from \cite{2017Natur.542..456G} and they are listed in Table~\ref{tab:trappist1}.

The results from the simulation are shown in Fig.~\ref{fig:trappist1}. Given our initial conditions, the system evolves with 3:2 resonances between planets 3-4, 4-5 and 6-7, a 4:3 resonance between planets 5-6, and a 8:3 resonance between planets 1-2. A more extensive and detailed study of this system will be presented in an upcoming article.

\section{Conclusions}

Posidonius is a fast and accurate open source N-body simulator for planetary systems that takes into account tidal effects (following the recipes described in \citealt{2015A&A...583A.116B}) and adds new updated stellar evolution models. Its high performance and memory safety are ensured thanks to the use of Rust as programming language. To create simulation cases with initial conditions, Posidonius offers a user-friendly scripting interface based on Python. The code was assessed by successfully reproducing a selection of \mbox{Mercury-T} simulations presented in \cite{bolmo2013} and \cite{2015A&A...583A.116B}. Based on these results, we showed that:

\begin{itemize}
    \item Posidonius is more than six times faster than \mbox{Mercury-T};
    \item It conserves the total angular momentum of the system one order of magnitude better than \mbox{Mercury-T};
    \item It conserves the spin to rotational-flattening evolution five orders of magnitude better than \mbox{Mercury-T}.
\end{itemize}

The simulations can be stopped and resumed at any moment thanks to the automatic creation of recovery snapshots. Additionally, any simulation can be reproduced by just re-using the case definition scripts that contain the initial conditions, the JSON file that the script generates or any intermediary binary snapshot. All the results presented in this work make Posidonius a powerful tool for N-body simulations of planetary systems and, particularly, for the study of tidal effects.

\section*{Acknowledgments}
{This work would not have been possible without the support of Dr. Laurent Eyer (University of Geneva). The authors thank Christophe Cossou for his Python routines to explore resonances. E.B. acknowledges funding by the European Research Council through ERC grant SPIRE 647383. This research has made use of NASA's Astrophysics Data System. All the software and technology used to build Posidonius were provided by the open source community.}

\bibliographystyle{ewass_ss4proc}
\bibliography{posidonius.bib}

\begin{thebibliography}{23}
\providecommand{\natexlab}[1]{#1}

\bibitem[\protect\astroncite{{Anderson}
  \emph{et~al.}}{1975}]{1975ApJ...200..221A}
{Anderson}, J.~D., {Esposito}, P.~B., {Martin}, W., {Thornton}, C.~L., \&
  {Muhleman}, D.~O. 1975, \apj, 200, 221.

\bibitem[\protect\astroncite{{Baraffe}
  \emph{et~al.}}{1998}]{1998A&A...337..403B}
{Baraffe}, I., {Chabrier}, G., {Allard}, F., \& {Hauschildt}, P.~H. 1998, \aap,
  337, 403.

\bibitem[\protect\astroncite{{Blanco-Cuaresma} \&
  {Bolmont}}{2017}]{2017IAUS..325..341B}
{Blanco-Cuaresma}, S. \& {Bolmont}, E. 2017, In \emph{Astroinformatics}, edited
  by M.~{Brescia}, S.~G. {Djorgovski}, E.~D. {Feigelson}, G.~{Longo}, \&
  S.~{Cavuoti}, \emph{IAU Symposium}, vol. 325, pp. 341--344.

\bibitem[\protect\astroncite{Bolmont}{2013}]{bolmo2013}
Bolmont, E. 2013, \emph{Evolution et habitabilit\'e de syst\`emes plan\'etaires
  autour d'\'etoiles de faible masse et de naines brunes}.
\newblock Ph.D. thesis.
\newblock Th\`ese de doctorat dirig\'ee par Raymond, Sean N. et Selsis, Franck
  Astrophysique Bordeaux 1 2013.

\bibitem[\protect\astroncite{{Bolmont}
  \emph{et~al.}}{2017}]{2017A&A...604A.113B}
{Bolmont}, E., {Gallet}, F., {Mathis}, S., {Charbonnel}, C., {Amard}, L.,
  \emph{et~al.} 2017, \aap, 604, A113.

\bibitem[\protect\astroncite{{Bolmont} \& {Mathis}}{2016}]{2016CeMDA.126..275B}
{Bolmont}, E. \& {Mathis}, S. 2016, Celestial Mechanics and Dynamical
  Astronomy, 126, 275.

\bibitem[\protect\astroncite{{Bolmont}
  \emph{et~al.}}{2011}]{2011A&A...535A..94B}
{Bolmont}, E., {Raymond}, S.~N., \& {Leconte}, J. 2011, \aap, 535, A94.

\bibitem[\protect\astroncite{{Bolmont}
  \emph{et~al.}}{2015}]{2015A&A...583A.116B}
{Bolmont}, E., {Raymond}, S.~N., {Leconte}, J., {Hersant}, F., \& {Correia},
  A.~C.~M. 2015, \aap, 583, A116.

\bibitem[\protect\astroncite{Bulirsch}{2013}]{Stoer2013}
Bulirsch, J.~S. 2013, \emph{Introduction to numerical analysis}, vol.~12.

\bibitem[\protect\astroncite{{Chambers}}{1999}]{1999MNRAS.304..793C}
{Chambers}, J.~E. 1999, \mnras, 304, 793.

\bibitem[\protect\astroncite{{de Souza Torres} \&
  {Anderson}}{2008}]{2008arXiv0808.0483D}
{de Souza Torres}, K. \& {Anderson}, D.~R. 2008, ArXiv e-prints.

\bibitem[\protect\astroncite{{Gallet}
  \emph{et~al.}}{2017}]{2017A&A...604A.112G}
{Gallet}, F., {Bolmont}, E., {Mathis}, S., {Charbonnel}, C., \& {Amard}, L.
  2017, \aap, 604, A112.

\bibitem[\protect\astroncite{{Gillon}
  \emph{et~al.}}{2016}]{2016Natur.533..221G}
{Gillon}, M., {Jehin}, E., {Lederer}, S.~M., {Delrez}, L., {de Wit}, J.,
  \emph{et~al.} 2016, \nat, 533, 221.

\bibitem[\protect\astroncite{{Gillon}
  \emph{et~al.}}{2017}]{2017Natur.542..456G}
{Gillon}, M., {Triaud}, A.~H.~M.~J., {Demory}, B.-O., {Jehin}, E., {Agol}, E.,
  \emph{et~al.} 2017, \nat, 542, 456.

\bibitem[\protect\astroncite{{Hernandez} \&
  {Dehnen}}{2017}]{2017MNRAS.468.2614H}
{Hernandez}, D.~M. \& {Dehnen}, W. 2017, \mnras, 468, 2614.

\bibitem[\protect\astroncite{{Jehin} \emph{et~al.}}{2011}]{2011Msngr.145....2J}
{Jehin}, E., {Gillon}, M., {Queloz}, D., {Magain}, P., {Manfroid}, J.,
  \emph{et~al.} 2011, The Messenger, 145, 2.

\bibitem[\protect\astroncite{{Leconte} \&
  {Chabrier}}{2013}]{2013NatGe...6..347L}
{Leconte}, J. \& {Chabrier}, G. 2013, Nature Geoscience, 6, 347.

\bibitem[\protect\astroncite{{Leconte}
  \emph{et~al.}}{2011}]{2011A&A...528A..41L}
{Leconte}, J., {Lai}, D., \& {Chabrier}, G. 2011, \aap, 528, A41.

\bibitem[\protect\astroncite{{Newhall}
  \emph{et~al.}}{1983}]{1983A&A...125..150N}
{Newhall}, X.~X., {Standish}, E.~M., \& {Williams}, J.~G. 1983, \aap, 125, 150.

\bibitem[\protect\astroncite{{Rein} \& {Liu}}{2012}]{2012A&A...537A.128R}
{Rein}, H. \& {Liu}, S.-F. 2012, \aap, 537, A128.

\bibitem[\protect\astroncite{{Rein} \& {Tamayo}}{2015}]{2015MNRAS.452..376R}
{Rein}, H. \& {Tamayo}, D. 2015, \mnras, 452, 376.

\bibitem[\protect\astroncite{{Rein} \& {Tamayo}}{2017}]{2017MNRAS.467.2377R}
{Rein}, H. \& {Tamayo}, D. 2017, \mnras, 467, 2377.

\bibitem[\protect\astroncite{{Siess} \emph{et~al.}}{2000}]{2000A&A...358..593S}
{Siess}, L., {Dufour}, E., \& {Forestini}, M. 2000, \aap, 358, 593.

\end{thebibliography}

\end{document}